\documentclass [prl, twocolumn, showpacs,showkeys, floatfix]{revtex4}
\usepackage{graphicx}
\usepackage{dcolumn}
\usepackage{bm}% bold math
\usepackage{amssymb}

\begin{document}

\title
 {Liquid-liquid phase separation of a surfactant-solubilized membrane protein}

\author{Roberto Piazza}
\email{roberto.piazza@polimi.it}
\author{Matteo Pierno}
\author{Emanuele Vignati}
\affiliation{ INFM - Politecnico di Milano, Dipartimento di Ingegneria Nucleare, via
Ponzio 34/3, I-20133 Milano (Italy)}
\author{Giovanni Venturoli}
\author{Francesco Francia}
\affiliation{ INFM -Dipartimento di Biologia, Universit\`a di Bologna, via Irnerio 42,
I-40126 Bologna (Italy)\\}
\author{Antonia Mallardi}
\author{Gerardo Palazzo}
\affiliation{IPCF-CNR sezione di Bari, and Dipartimento di Chimica, Universit\`a di Bari,
via Orabona 4, I-70126 Bari (Italy)}


\begin {abstract}
The phase behavior of membrane proteins stems from a complex synergy with the amphiphilic
molecules required for their solubilization. We show that ionization of a pH-sensitive
surfactant, LDAO, bound to a bacterial photosynthetic protein, the Reaction Center (RC),
leads in a narrow pH range to protein liquid-liquid phase separation in surprisingly
stable `droplets', forerunning reversible aggregation at lower pH. Phase segregation is
promoted by increasing temperature and hindered by adding salt.  RC light-absorption and
photoinduced electron cycle are moreover strongly affected by phase segregation.

\end{abstract}

\maketitle

Recent studies of protein solutions shed new light on soft-matter physics by blowing up
the conceivable panorama of complex-fluid interactions and phase behavior \cite{pia}. So
far, general attention has been mainly turned to \emph{soluble} proteins, performing
basic enzymatic or transport tasks. The very basic purpose of cell membranes, creating
compartments for life, would however be pointless without \emph{transmembrane} proteins
carrying on all primary exchange functions as specific ion channels, receptors for
extracellular signals, energy harvesters, `weavers' of the membrane texture driving cell
recognition, and linkers in cell adhesion \cite{alb}. Our knowledge of membrane proteins
is unfortunately rather poor, even at the structural level. Obtaining membrane protein
crystals is indeed extremely hard, and crystallization protocols have so far been
successful only for a very limited number of proteins \cite{mic}. The hampering fact is
that membrane proteins, displaying large exposed hydrophobic regions often associated
with lipids, are essentially insoluble in water. Stable solutions can only be obtained by
exploiting the solubilizing properties of specific detergents used for extracting them
from the supporting cell membrane. Surfactants are therefore necessary `chaperons' in
order to bring membrane protein in solutions, and self-assembly phenomena unavoidably
coexist with, and strongly influence, membrane-protein solution behavior. The phase
behavior of membrane proteins stems therefore from a complex synergy between surfactant
supramolecular aggregation and protein-surfactant specific interactions. So far, however,
studies of the solution properties of membrane proteins are totally lacking.

The Reaction Center (RC) is a 100 kDa bacterial pigment-protein complex spanning the
intracytoplasmic membrane and accomplishing the primary events of energy transduction by
promoting light-induced charge separation across the membrane \cite{feh}. RC, which has
been the first membrane protein to be crystallized, can be extracted from bacterial
membranes by using lauryldymethylamino-N-oxide (LDAO), a surfactant acting as a very
efficient solubiliser due to the good matching between its spontaneous curvature and the
packing constraints imposed by RC structure. In this Letter we show that the interplay
between the ionization state of LDAO, tuned by the solution pH, and the packing
requirements for efficient solubilization of the Reaction Center induces complex
association effects, leading in a narrow pH range to segregation of the
protein-surfactant complexes into mesoscopic `droplets', with a typical size of the order
of few $\mu$m and a relatively narrow size distribution. This micro-phase segregation
shows many features resembling a liquid-liquid phase separation. Furthermore, at variance
with what is found for most spontaneous aggregation phenomena in complex fluids, phase
segregation of RC/LDAO complexes is fully hindered by screening of the electrostatic
interactions with the addition of salt. Finally, confinement into droplets has noticeable
effects on RC photochemical reactivity for what concerns both protein absorption spectrum
and dynamics of ground-state recovery after excitation. Besides describing an uncommon
spontaneous re-organization of a complex fluid involving formation of structures far
larger than the basic constituents, these findings, stressing the primary influence of
the surrounding environment on RC biochemical activity, may therefore have impact on our
understanding of membrane protein functionality within the `macromolecular crowding'
constituting the biological cell.

Reaction Center was isolated and purified from the purple bacterium \textit{Rhodobacter
sphaeroides} R-26 according to Gray et al. \cite{gra}. We first discuss the qualitative
phase behavior of RC/LDAO solutions observed in moderate acid conditions, at room
temperature, and in absence of added salt. From the original stock solution, samples were
prepared at RC concentration $c= 4.4~\mathrm{\mu M}$ in presence of 0.87 mM LDAO (value
below the surfactant critical micellar concentration $c.m.c\simeq$~2 mM), and the
solution pH tuned in the range $5<pH<8$ by adding HCl to 10 mM imidazole buffers. For
$\mathrm{pH} > 7$ solutions look optically transparent and scatter light very weakly. By
increasing acidity, a progressive growth of turbidity associated to light scattered at
small angles is observed, until at $\mathrm{pH} < 6$ fast aggregation of the solute,
which precipitates as an amorphous powder, takes place. Aggregation is however almost
fully reversible, as confirmed by the rapid dissolution of most of the aggregates
observed by titrating the solution back to pH = 8. For $\mathrm{pH} \simeq 5$
precipitation kinetics becomes slower, but aggregates are harder to break up. Solution
behavior around pH=6.5 is however very peculiar. In this region, samples persist in a
strongly turbid state for days (with typical extinction coefficients of the order of
$3-4~\mathrm{cm^{-1}}$), with little solute precipitation or sedimentation. Severe
problems to obtain homogeneous RC/LDAO solutions around pH = 6.5 have indeed been
previously reported \cite{gas}, but no detailed scrutiny of the effect has so far been
performed.
\begin {figure}[h]
\includegraphics [clip, width =8.4 cm]{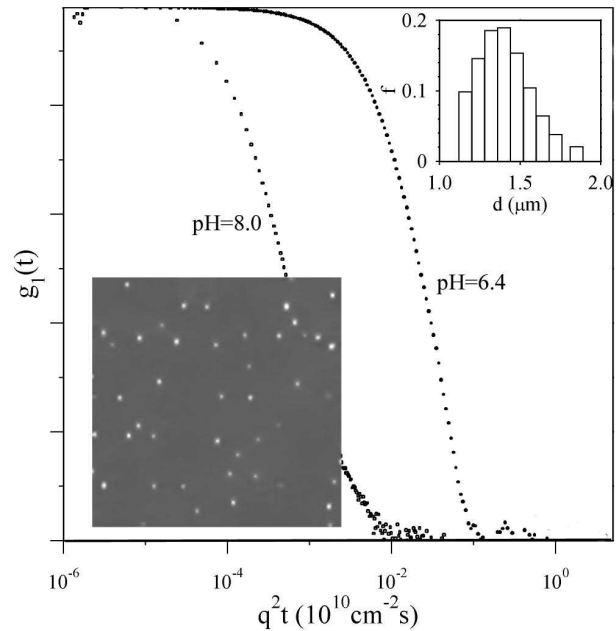}
 \caption{\label {f.1} Main body: DLS field correlation functions for RC/LDAO solutions at T=25~$^\circ$C.
 Upper right inset: DLS particle size distribution at pH=6.5. Lower
 left inset: Phase-contrast image of phase-separated droplets at pH = 6.5. Field of view is
 $80 \times 80~\mathrm{\mu m^2}$.}
\end {figure}
RC absorbance, which is strong in the blue-violet and near-infrared regions, is low
within a moderately large visible spectral range, allowing for dynamic light scattering
(DLS) measurements. By progressively changing the pH of dilute RC/LDAO solutions from 8
to 6.5, the scattered intensity not only grows by order of magnitudes, but also becomes
strongly forward-peaked. DLS measurements concurrently show a dramatic slowing down of
translational diffusion. Time-correlation functions at pH = 8 (Fig.~\ref{f.1}, main body)
give a hydrodynamic radius $R_H\approx 5$~nm for the RC/LDAO complex (yet, we
unexpectedly detected the presence of a very small quantity of objects with a size around
20-50 nm, persisting even after extensive close-loop filtering). Conversely, DLS
measurements at pH=6.5 suggest the presence of much larger objects, with a typical size
more than two order of magnitude larger. The calculated particle size distribution shown
in the upper right inset, gives an average diameter $d\approx 1.4~\mathrm{\mu m}$ with a
standard deviation of about 40\%. Direct visualization, made by sealing samples into
rectangular capillaries and using phase-contrast microscopy, show that strong scattering
is due to a relatively large number of liquid-like droplets (Fig.1, lower left inset),
with a typical diameter around $1-2~\mathrm{\mu m}$. By counting the number of droplets
within the depth of field of a 60X, 0.7 NA objective, we estimate that samples at pH =
6.5 with $1\mu$M RC, 0.87 mM LDAO , approximately contain a volume fraction $\Phi
=2\times10^{-3}$ of dispersed droplets. When filtered through a $0.2~\mathrm{\mu m}$ low
protein-binding membrane, samples where observed to clarify considerably, meaning that
most of the droplets are blocked by the filter. Indeed, DLS measurements of filtered
samples essentially coincides with those obtained at pH=8. By comparing absorption
spectra of filtered samples at pH=6.5 and pH=8, it is possible to conclude that at least
95\% of the original protein amount is confined within the droplets. On the basis of the
previous estimate of $\Phi$, this means that RC concentration within the droplets may
reach up to 2~mM. LDAO concentration within the separated phase could not be easily
measured, but since in this pH range LDAO associates to RC in a 300/400 molar ratio
\cite{gas}, it should range between 15 to 20 \%. Such a high value is consistent with the
observation that, by letting samples dry at very slow rate under crossed polarizers,
droplets start to display, after limited volume shrinkage, liquid crystal textures
observed for pure LDAO at $c \gtrsim 40\%$ \cite{fuk}.

\begin {figure}[h]
\includegraphics [clip, width =8.4 cm]{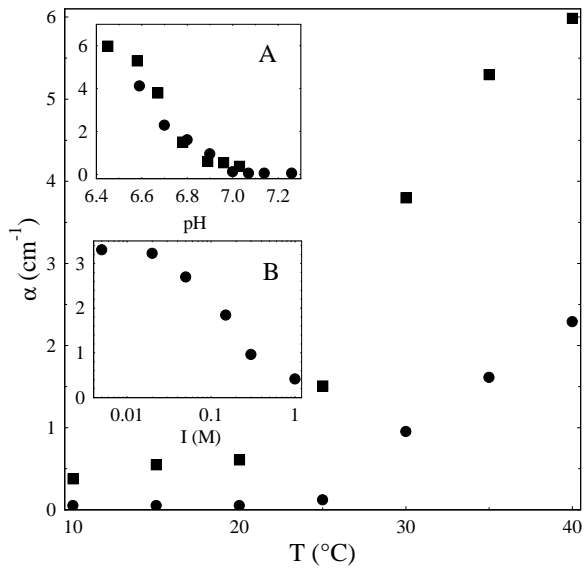}
 \caption{\label {f.2} Temperature dependence of the scattering extinction coefficient  $\alpha$ @
$\lambda = 633$~nm for $4.4\mu$M RC solutions in presence of 0.87 mM  LDAO at pH=7.0
(dots) and 6.75 (squares). Same data are plotted in Inset A as a function of measured
pH(T). Inset B: Ionic strength dependence of $\alpha$ for RC solutions at pH=6.5 and T=20
$^\circ$C.}
\end {figure}

We have checked whether the observed behavior depends on temperature. Fig. \ref{f.2}
shows that the scattering extinction coefficient $\alpha$, measured at pH values close to
the onset of the `anomalous' region, strongly increases with T. For instance, $4.4~\mu$M
RC solutions at pH=7, looking almost transparent at room temperature, show an extinction
coefficient $\alpha \approx 1 \mathrm{cm^{-1}}$ at T=$30^{\circ}$~C. Therefore, droplet
formation could be possibly associated to the onset of an inverted liquid-liquid phase
separation. Yet, little droplet coalescence and almost no macroscopic phase segregation
are observed over long time-scales, due to very good density-matching between the two
phases. While RC has typical protein density $\rho_P \approx 1.35$~g/l, LDAO is indeed
lighter than water ($\rho_S \approx 0.9$~g/l), so that, by using the previous estimates
for RC and LDAO concentration, droplet density is found to be around 1.03~g/l. A final
and very peculiar aspect of RC/LDAO liquid-liquid phase separation is that it is
\emph{fully hindered} by increasing the solution ionic strength with the addition of KCl.
Inset B of Fig.~\ref{f.2} shows indeed that strong turbidity enhancement at pH = 6.5 is
confined to moderately low electrolyte concentration, and progressively vanishes by
increasing the ionic strength up to 1 M.

The observed phase behavior for RC/LDAO solutions is strongly correlated with the
ionization properties of the surfactant. LDAO is indeed a pH-sensitive amphiphile, which
is nonionic at neutral and basic pH and becomes increasingly protonated (cationic) in
acid conditions (reported pK = 5.0 \cite{mae}). Charging of LDAO may be expected to have
strong effects on RC/LDAO association for two concurrent reasons. First of all, since RC
net charge in this pH range is negative, due to excess of acidic aminoacids in the
protein hydrophilic caps, mutual neutralization of the protein and surfactant opposite
charges may be expected to deplete the `surfactant belt' around the protein hydrophobic
region and decrease complex solubility. The charge-neutrality point of RC/LDAO complex
has indeed been reported to be $\mathrm{pI}\approx 6.1$ \cite{duc}, quite close to the
phase-separation region. At the same time, LDAO spontaneous curvature is larger when the
surfactant head group is charged, leading to worse structural matching. We point out that
similar effects due to surfactant protonation have been clearly detected by Mel'nikova
and Lindman studying DNA condensation in presence of LDAO \cite{lin}. As we have seen,
RC/LDAO liquid-liquid phase separation takes place by increasing temperature, at variance
with simple binary mixtures, but similarly, for instance, to what happens (at much higher
protein concentration) for normal and sickle-cell hemoglobin (Hb) \cite{vek}. It is
interesting to notice that, also for sickle-cell Hb, liquid-liquid separation `foreruns'
much more extensive protein association, in form of Hb polymerization. Inverted
miscibility gaps necessary call for temperature-dependent interactions, but is generally
hard to single out and quantify specific temperature effects on interparticle forces. The
situation may be simpler for RC/LDAO, since temperature changes definitely influence
hydrogen-ion activity of the buffer, modifying pH and therefore surfactant ionization
(for imidazole buffer $d(pK_a)/dT \approx -0.017~^{\circ}\mathrm{C^{-1}}$). Inset~A in
Fig~\ref{f.2} shows indeed that the $\alpha(T)$ for samples with different pH at T =
25~$^\circ$C collapse on a single curve when plotted versus the actual pH(T), directly
measured with temperature-compensated electrodes. Our hypothesis of a mutual LDAO/RC
charge-neutralization process, driving reduced RC solvation and eventually leading to
phase-separation, is further supported by the ionic strength dependence of the effect,
since salt addition would screen RC/LDAO opposite charge attraction.

We discuss now the effect of phase segregation on RC photochemistry, giving a rough
picture of RC electron-transfer cycle. At variance with green plants or algae, energy
fruition by photosynthetic bacteria is not based on direct water splitting, but rather
depends on simultaneous reduction of quinones, small hydrophobic molecules fully mobile
within the cell membrane. Within the RC, a bacteriochlorophyll dimer (D) acts as primary
elctron donor. Following photon absorption, it delivers an electron in sequence to a
couple of quinones ($\mathrm{Q_A}$ and $\mathrm{Q_B}$), generating a $\mathrm{D^+Q_B^-}$
state. \textit{In vivo}, $\mathrm{D^+}$ is reduced by a small soluble protein, Citochrome
$\mathrm{C_2}$ (Cyt). Further double electron-transfer leads to formation of
$\mathrm{Q_BH_2}$, which freely diffuses to another membrane protein complex (bc1) where
it is finally set back to the initial charge state. This cyclic electron-transfer chain
creates the proton concentration gradient across the membrane needed to drive ATP
synthesis. \textit{In vitro}, the light-induced $\mathrm{D^+Q_B^-}$ state undergoes slow
charge recombination, which can be probed by exciting the state using a short
light-flash, and following the time evolution of specific spectral signatures of the
charge separated state \cite{mal}. In presence of exogenous Cyt, however, which is a very
efficient electron donor for $\mathrm{D^+}$, $\mathrm{D^+Q_B^-}$ recombination is
quenched, and spectral evolution becomes a straight probe of cytochrome reduction. We
point out that Cyt interacts primarily with the hydrophilic region of the RC, testing
therefore the local surrounding of this specific moiety.
\begin {figure}[t]
\includegraphics [width =8.4 cm]{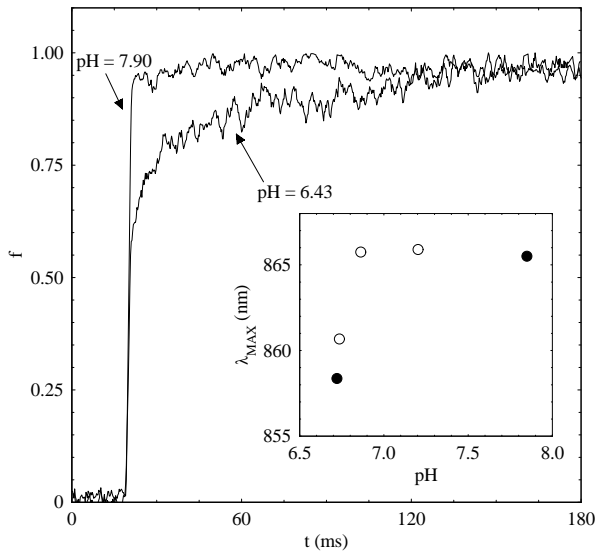}
 \caption{\label {f.3} Time-dependence of the fraction $f$ of oxidated cytochrome
 after excitation with a 20 ms flash. Inset: Shift of the maximum of 870 nm Bacteriochlrophyll dimer
 spectral band as a function of pH for RC/LDAO solutions in imidazole
 ($\bullet$) and MES ($\circ$) buffers.}
\end {figure}

A very schematic recollection of our findings is the following. By decreasing the pH
below 7, a relevant frequency shift (more than 10 nm) and intensity decrease of the
spectral band at 870 nm, due to a specific absorption by the bacteriochlorophyll dimer D,
is observed. Band intensity becomes negligible for $pH \leq 6.5$ (Fig.~\ref{f.3}, inset).
Band shift takes place regardless of the solubilizing buffer (imidazole, MES, or Tris),
and is not therefore due to specific buffer effects.  Moreover, Cyt oxidation, which at
pH $> 7$ is extremely fast, proceeds at a slower pace at pH = 6.5, reaching full
completion only after few tens of ms. The RC-Cyt redox reaction is extremely fast when
the two proteins are bound in a pre-existing complex (which is therefore the case for
$\mathrm{pH}>7$), while it is collision-limited, and therefore much slower, if Cyt is
free. Overall kinetics at pH = 6.5 can be quantitatively described by assuming that about
half of the RCs undergo reduction by collisional reaction with Cyt. Although we must
defer detailed discussion of phase-separation effects on RC photochemistry to a future
publication, we regard these findings as consistent with the general picture we have
suggested. Quite similar blue-shift and intensity decrease of the 870 nm band is indeed
observed when RC solutions are prepared at LDAO concentration lower than what needed for
full RC `coating' \cite{gas}, suggesting that phase-separation is driven by increasing
hydrophobicity due to surfactant desorption from RC non-polar regions. Cyt oxidation
kinetics becomes to a large extent collisional because of competitive binding of LDAO to
RC polar caps, leading to charge-neutralization, and limiting at the same time the
occurrence of RC/Cyt pre-formed complexes. Due to the presumably high viscosity of the
surrounding concentrated LDAO solution, diffusion-limited Cyt oxidation is moreover
particularly slow. We finally observed that, at pH = 6.5, charge recombination of the
$\mathrm{D^+Q_B^-}$ state markedly slows down, possibly reflecting an altered stability
of $\mathrm{Q_B^-}$ in the phase-separated state. Consistently with turbidity, DLS, and
direct visualization results, all these `anomalous' kinetic and spectral effects vanish
in presence of a sufficient amount of added KCl.

In conclusion, we have seen that RC/LDAO solutions may undergo micro-phase segregation
processes which are totally absent for pure LDAO solutions, take place at very low
protein concentration, strongly resemble an inverted liquid-liquid phase separation,
forerun extensive (but reversible) aggregation, and disappear by increasing electrostatic
screening. The origin of such complex cooperative effects may be likely traced down to
RC/LDAO mutual charge neutralization effects, leading to less efficient screening of
protein hydrophobic regions and to effective attractive interparticle interactions.
Furthermore, phase-segregation effects on RC photochemical kinetics  support the proposed
mechanism, and suggest that full intelligence of membrane protein operation has to make
allowance for protein specific association state.

\begin{acknowledgements}

\end{acknowledgements}

\end{document}